%% file: 0_brainnet.tex
\documentclass[sigconf, authorversion]{acmart}

\AtBeginDocument{%
  \providecommand\BibTeX{{%
    \normalfont B\kern-0.5em{\scshape i\kern-0.25em b}\kern-0.8em\TeX}}}

\copyrightyear{2022}
\acmYear{2022}
\setcopyright{acmlicensed}\acmConference[KDD '22]{Proceedings of the 28th ACM SIGKDD Conference on Knowledge Discovery and Data Mining}{August 14--18, 2022}{Washington, DC, USA}
\acmBooktitle{Proceedings of the 28th ACM SIGKDD Conference on Knowledge Discovery and Data Mining (KDD '22), August 14--18, 2022, Washington, DC, USA}
\acmPrice{15.00}
\acmDOI{10.1145/3534678.3539178}
\acmISBN{978-1-4503-9385-0/22/08}

\usepackage{textcomp}
\usepackage{verbatim}
\usepackage{amsfonts}
\usepackage{amsmath}
\usepackage{array}
\usepackage{subfigure}
\usepackage{caption}
\usepackage{balance}
\usepackage{multicol}
\usepackage{multirow}
\usepackage{color}
\usepackage{epsfig}
\usepackage{natbib}
\usepackage{xspace}
\usepackage{booktabs}
\usepackage{bm}
\usepackage{enumitem}
\usepackage{diagbox}
\usepackage{url}
\usepackage{geometry}
\usepackage{changepage}
\usepackage{textcomp}
\usepackage{graphicx}
\usepackage{color}
\usepackage{bm}
  
\usepackage{amssymb}
\usepackage{algorithm}
\usepackage{algorithmic}
\usepackage{threeparttable}
\usepackage{mathrsfs}

\newcommand{\model}{\textit{BrainNet}\xspace}

\newtheorem{definition}{Definition}

\newcommand{\vpara}[1]{\vspace{0.05in}\noindent\textbf{#1 }}
\newcommand{\hide}[1]{} 

\begin{document}

\title[BrainNet: Epileptic Wave Detection from SEEG with Hierarchical Graph Diffusion Learning]{BrainNet: Epileptic Wave Detection from SEEG \\ with Hierarchical Graph Diffusion Learning}

\author{Junru Chen}
\authornote{Both authors contributed equally to this research.}
\affiliation{%
  \institution{Zhejiang University}
  \country{}
}
\email{jrchen\_cali@zju.edu.cn}

\author{Yang Yang}
\authornotemark[1]
\affiliation{%
  \institution{Zhejiang University}
  \country{}
}
\email{yangya@zju.edu.cn}
\authornote{Corresponding author.}

\author{Tao Yu}
\affiliation{%
  \institution{Zhejiang University}
  \country{}
}
\email{yutaoyt@zju.edu.cn}

\author{Yingying Fan}
\affiliation{%
  \institution{Zhejiang University}
  \country{}
}
\email{yyingfan@outlook.com}

\author{Xiaolong Mo}
\affiliation{%
  \institution{Neuroechos Medical}
  \country{}
}
\email{Xiaolong.mo@neurox.cn}

\author{Carl Yang}
\affiliation{%
  \institution{Emory University}
  \country{}
}
\email{j.carlyang@emory.edu}

\renewcommand{\shortauthors}{Junru Chen et al.}

\input{1_abstract}



\maketitle

\input{2_introduction}
\input{3_problem}
\input{4_model}
\input{5_experiment}
\input{6_related}
\input{7_conclusion}

\bibliographystyle{ACM-Reference-Format}
\bibliography{reference}

\clearpage
\appendix
\input{8_appendix}

\end{document}

%% file: 1_abstract.tex
\begin{abstract}

Epilepsy is one of the most serious neurological diseases, affecting 1-2$\%$ of the world's population. 
The diagnosis of epilepsy depends heavily on the recognition of \textit{epileptic waves}, \emph{i.e.}, disordered electrical brainwave activity in the patient's brain. 
Existing works have begun to employ machine learning models to detect epileptic waves via \textit{cortical electroencephalogram} (EEG), which refers to brain data obtained from a noninvasive examination performed on the patient's scalp surface to record electrical activity in the brain.
However, the recently developed \textit{stereoelectrocorticography} (SEEG) method provides information in stereo that is more precise than conventional EEG, and has been broadly applied in clinical practice. 
Therefore, in this paper, we propose the first data-driven study to detect epileptic waves in a real-world SEEG dataset.
While offering new opportunities, SEEG also poses several challenges.
In clinical practice, epileptic wave activities are considered to propagate between different regions in the brain. 
These propagation paths, also known as the \textit{epileptogenic network}, are deemed to be a key factor in the context of epilepsy surgery. However, the question of how to extract an exact epileptogenic network for each patient remains an open problem in the field of neuroscience. 
Moreover, the nature of epileptic waves and SEEG data inevitably leads to extremely imbalanced labels and severe noise. 
To address these challenges, we propose a novel model (\model) that jointly learns the dynamic diffusion graphs and models the brain wave diffusion patterns. 
In addition, our model effectively aids in resisting label imbalance and severe noise by employing several self-supervised learning tasks and a hierarchical framework. 
By experimenting with the extensive real SEEG dataset obtained from multiple patients, we find that \model outperforms several latest state-of-the-art baselines derived from time-series analysis. 

\end{abstract}

%% file: 2_introduction.tex
\section{Introduction}~\label{sec:intro}

\begin{figure}[t]
  \centering
  \includegraphics[width=\linewidth]{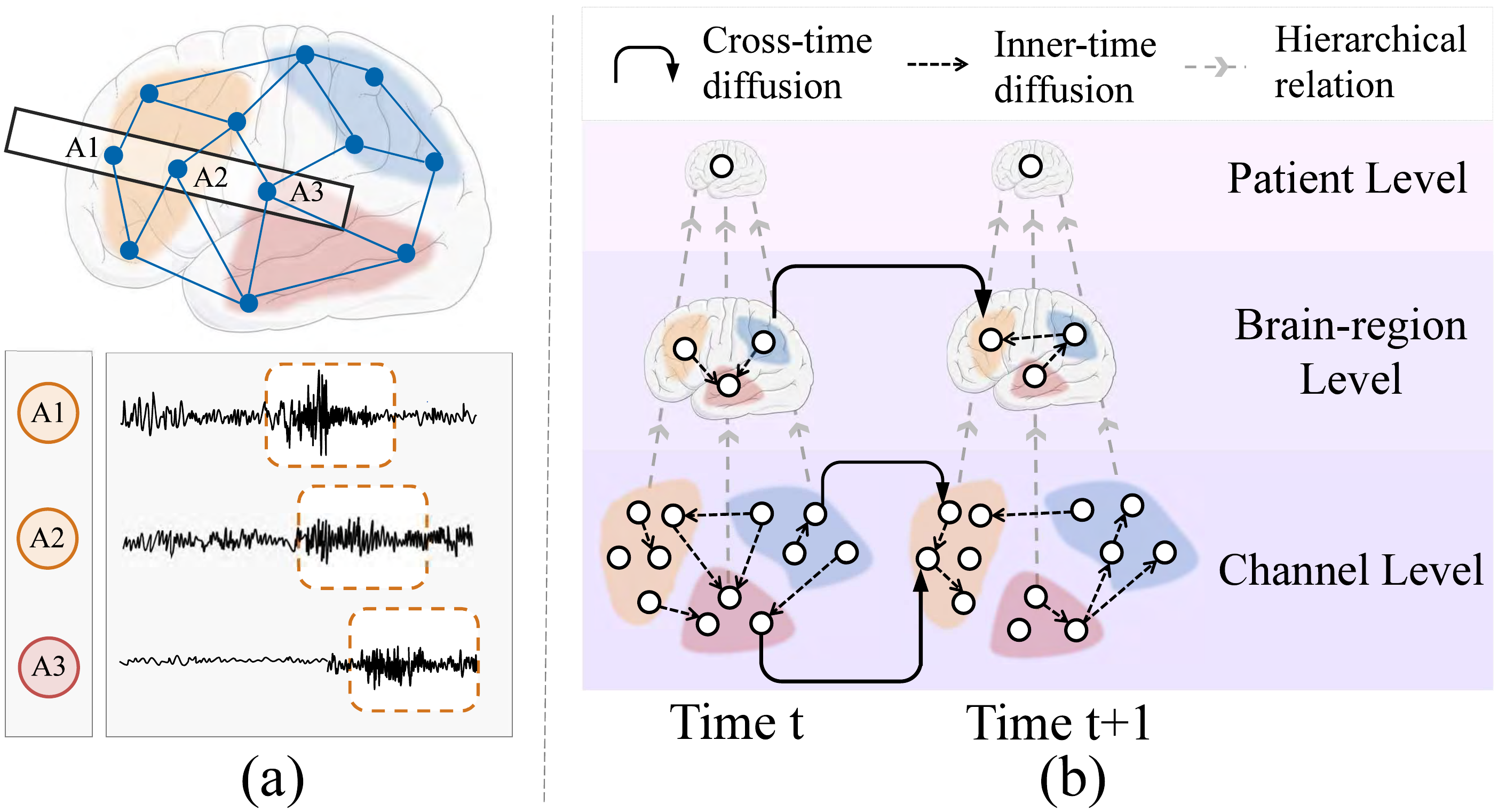}
  \caption{
  Illustration of epileptic wave detection. 
  \small 
  (a) An electrode with three channels (A1, A2, and A3) is inserted into the brain to collect SEEG signals across two regions of a patient's brain. 
  (b) Our solution is to jointly learn the process by which epileptic seizures diffuse and detect epileptic waves, which are marked by yellow squares among the SEEG signals in (a), on multiple levels. 
  \normalsize 
  }
  \label{pic:story}
\end{figure}

\vpara{Background.}
\textit{Epilepsy}, which is one of the most common serious neurological diseases, is characterized by abnormal neurophysiological activity, leading to epileptic seizures or abnormal behavior.
Today, in 2021, there are more than 65 million epilepsy patients globally, approximately one-third of whom are medication-resistant~\cite{ni2021pathogenic}. 
In other words, medication will not be effective for these patients, and surgical removal of the area of the brain involved in seizures is considered the only effective treatment. 

To measure the \textit{seizure onset zone} (SOZ) or so-called \textit{epileptogenic foci}, and guide epilepsy surgery, it is necessary to record the electrical activity of the patient’s brain.
There are two main types of such electrophysiological monitoring methods: EEG and SEEG.
The former is noninvasive, while the latter is invasive (\emph{i.e.}, requires electrodes to be inserted into the brain), and thus contains more stereo information. 
For example, when the SOZ is located in the deeper structures of the brain (such as the hippocampus or the insula), or when the \textit{laterality of seizures} \cite{khuvis2021intracranial} is unknown, non-invasive testing will fail to pinpoint the exact seizure focus, in which case the SEEG approach is necessary.

\vpara{Problem.} 
To facilitate the development of epilepsy treatment, we collect a real-world SEEG dataset, which is made up of high-frequency and multi-channel SEEG signals obtained from multiple epilepsy patients (each patient has a 77GB record of 53 hours on average) in a specific first-class hospital. 
Based on the dataset, we further propose to automatically detect epileptic waves. 
Besides, the fact that epileptic wave activities propagate among different brain regions in clinical practice inspires us to further study the underlying \textit{epileptogenic network}~\cite{worrell2008high}, which characterizes the relations among the brain regions involved in the production and propagation of epileptic activities. 
It is deemed to be a key factor in the context of epilepsy surgery, but how to extract an exact epileptogenic network for each patient is still an open problem in neuroscience~\cite{bartolomei2017defining}. 

By the example illustrated in Figure~\ref{pic:story}, we further introduce the details of our problem. 
The top part of Figure~\ref{pic:story}(a) shows a human brain into which an electrode with three channels has been inserted from the upper left corner. 
Notably, in real clinical diagnosis, doctors will insert multiple electrodes (each with multiple channels) into the suspected epileptogenic areas of the brain, which vary across different patients and may span multiple brain regions. 
After inserting the electrode, the doctor can collect and monitor the patient's SEEG data; as the bottom part of Figure~\ref{pic:story}(a) shows, this can be regarded as a contiguous multi-channel time series. 

Given the SEEG data, our target is to formulate an automatic data-driven method to pinpoint the time and location at which epileptic waves appear (marked by yellow squares in Figure~\ref{pic:story}(a)). 
It is worth mentioning that, in this work, we focus on epilepsy detection for individual patients. 
More specifically, given a particular patient, we aim to train a model based on his/her historical SEEG data, then utilize the model to identify epileptic waves in the current SEEG data, following existing works with similar settings on EEG data~\cite{kiral2018epileptic}.
An alternative solution would involve a model to handle all patients together. 
However, individual differences in epileptogenic foci result in deep subdural electrodes being utilized with different numbers and locations from person to person, which further leads to dramatic variation in the collected signals across patients. 
Before properly handling the signals of individual patients, it is hard to directly study the possible shared patterns and generalization of models across patients. 
Therefore, we focus on patient-specific models and predictions in this work as the first study on SEEG-based epilepsy detection.

\vpara{Challenges.}
Compared with existing works that study epileptic waves derived from the much simpler EEG data, there are several unique challenges for us in fully leveraging the SEEG data, which are caused by the nature of the data and the lack of understanding regarding the diffusion mechanisms of brain waves. 
\begin{itemize}[leftmargin=*]
    \item[1]Capturing the true \textit{epileptogenic network}~\cite{bartolomei2017defining}.
    The propagation of epileptic seizures cannot be observed directly and does not follow any known routines. 
    Indeed, an epileptic seizure will not diffuse in a manner consistent with the anatomical brain structure~\cite{de2007canonical}, while its propagation paths may dramatically change over time~\cite{bartolomei2017defining}. 
    Therefore, quantifying the dynamic diffusion graph (\emph{i.e.}, epileptogenic network) structure is challenging.
    \item[2]Handling imbalanced labels.
    SEEG often generates extremely large amounts of data. 
    This is partly due to its high frequency of data acquisition (mostly ranging from 256Hz to 1024Hz).
    Moreover, patients are monitored with electrodes for an average of 11 days, and sometimes for up to 33 days~\cite{mullin2016seeg}. However, the epileptic seizure process lasts only for tens of seconds among days of records. Consequently, the low epileptic wave rate leads to the imbalanced label issue.
    \item[3]Handling data noise.
    Due to inherent problems like flaws in electrode artifacts, mechanical noises, and the interference of epileptic interval waves, SEEG data is severely affected by noise.
\end{itemize}

\vpara{Solution.} 
To address the above challenges, in this paper, we propose a novel epileptic wave detection model, referred to as \model. 

To handle the first challenge, we aim to find out the underlying epileptogenic network tracking the process of epileptic wave diffusion across time.
To this end, \model adopts graph neural networks along with a structure learning algorithm in order to both learn and quantify the epileptic wave diffusion process. More specifically, as the duration of epileptic waves may be longer or shorter, \model learns two types of diffusion processes.
Longer epileptic waves naturally result in diffusion from one time segment to the next, which is called \textit{cross-time diffusion}, denoted by solid black lines in Figure~\ref{pic:story}(b). 
Meanwhile, within the same time segments of each channel, there also exists a diffusion process that occurs as the electric signals spread quickly.
The dotted black lines in Figure~\ref{pic:story}(b) represent the \textit{inner-time diffusion} described above.

For the second challenge, the \model adopts a self-supervised learning approach to overcome extremely imbalanced labels.
Specifically, we propose \textit{bidirectional contrastive predictive coding} (BCPC) to pre-train the representation of every segment in each channel. 
Compared with existing self-supervised algorithms, BCPC makes \model more capable of extracting bidirectional information by taking full advantage of the sufficient (unlabeled) SEEG data.

Finally, to make the model more robust to noise, we propose auxiliary learning tasks with a hierarchical framework. 
Inspired by the diagnosis process used by doctors, which usually considers information from different levels simultaneously—including channels (micro-level), brain regions (meso-level), and patients (macro-level)—to make more appropriate and accurate diagnoses, we propose to make \model further predict whether a particular brain region or patient will be epileptic at a specific time. 
More specifically, a sample in the brain-region/patient level is said to be normal if none of its corresponding channels/brain-regions are epileptic. 
By adopting this approach, the input sensitivity of lower levels will be weakened at higher levels. 
Intuitively, \model aggregates accurate information at higher levels, and in turn feeds it back to the lower levels, reducing the likelihood of inaccurate information being accumulated there.

To sum up, by utilizing all technical designs discussed above, \model possesses the ability to capture the dynamic diffusion process and enhance the accuracy of epileptic wave detection tasks.
The main contributions can be summarized as follows: 
\begin{itemize}[leftmargin=*]
    \item We are the first to formulate and study the epileptic wave detection problem using an automatic end-to-end data-driven method for SEEG data.
    \item We propose \model to jointly learn the dynamic diffusion graphs and model the brain wave diffusion patterns thereon to achieve accurate epileptic wave detection under conditions of imbalanced labels and severe noise.
    \item We conduct extensive experiments on a large-scale real-world SEEG dataset involving multiple patients. Our results validate the effectiveness of our model on epileptic wave detection, while case studies show its superiority to capture the diffusion process.
\end{itemize}

%% file: 3_problem.tex
\section{Problem Definition}~\label{sec:pre}

We take Figure~\ref{pic:story} as an example to illustrate our studied problem. 
In Figure~\ref{pic:story}(a), for an individual patient, we have the set of three brain regions\footnote{We define brain regions according to the automated anatomical labeling (AAL)~\cite{tzourio2002automated}, which is a digital atlas of the human brain and defines 116 different regions in total.} $\mathbf{B}$, represented by three colors.
All the electrode contacts (\emph{i.e.} blue solid nodes) consist of a set of channels $\mathbf{C}$.
Each channel $\mathbf{c} \in \mathbf{C}$ belongs to a unique brain region $b(\mathbf{c}) \in \mathbf{B}$, where $b(\cdot)$ is a function that maps a channel to its corresponding brain region, which is derived from domain knowledge.
For example, channels $A1$ and $A2$ both belong to the same orange brain region, while $A3$ is located in another red region. 
After locating the contactors, the machine begins to record the signals of every channel so as to collect the SEEG data—shown at the bottom of Figure~\ref{pic:story}(a)—as a multivariate time series $T=\{X_i\}_{i=1}^{|T|} \in \mathbb{R}^{|T|\times|\mathbf{C}|}$, where $X_i=\{x_{i,\mathbf{c}}\}_{\mathbf{c} \in \mathbf{C}} \in \mathbb{R}^{|\mathbf{C}|}$ represents a vector of the channel signals belonging to time point $i$. Moreover, $Y=\{Y_i\}_{i=1}^{|T|}$ denotes multivariate labels, where $Y_i=\{y_{i,\mathbf{c}}\}_{\mathbf{c} \in \mathbf{C}}$ are the labels of every channel at the $i$-th time point. Here, $y_{i,\mathbf{c}} \in \{0, 1\}$ indicates whether an epileptic seizure is occurring (\emph{i.e.} $y_{i, \mathbf{c}}=1$) or not (\emph{i.e.} $y_{i, \mathbf{c}}=0$). 
The positive sample ratio in practice is pretty low (around $0.003$). 

In line with existing works on time-series analysis \cite{bagnall2017great, schafer2015boss, lines2012shapelet}, we use a sliding window with length $k$ and stride $l$ to divide the raw time-series $T$ into smaller segments $S=\{S_t\}_{t=1}^{|S|}$ where $|S|=\lfloor (|T|-k)/l \rfloor+1$ is the number of segments. The annotation $S_t=\{s_{t,\mathbf{c}}\}_{\mathbf{c} \in \mathbf{C}} \in \mathbb{R}^{|\mathbf{C}| \times k}$ represents the segment with $|\mathbf{C}|$ channels and $s_{t,\mathbf{c}}=\{x_{l*(t-1)+i,\mathbf{c}}\}_{i=1}^{k} \in \mathbb{R}^{k}$ is the $t$-th time segment of one channel $\mathbf{c}$. Using the same division strategy, we can also divide the labels $Y$ into small segments $Y^S=\{Y_t^S\}_{t=1}^{|S|}$. If one segment includes a seizure point, we consider it to be a seizure segment; \emph{i.e.}, $y_{t,\mathbf{c}}^S=\max_{i=1}^{k}\{y_{l*(t-1)+i,\mathbf{c}}\}$. By means of basic time-series data segmentation, we formally define the studied problem as follows:
\begin{definition}[Epileptic Wave Detection]
  Given the historic segment set $\hat{S}$ and corresponding label set $Y^{\hat{S}}$ of an individual patient, 
  we aim to predict the labels $Y^{S}$ of the future segment set $S$:
  \begin{equation*}
      P(y^{S}_{t,\mathbf{c}}|S, \hat{S}, Y^{\hat{S}}), \qquad t=1,\dots,|S|, \quad \mathbf{c} \in \mathbf{C},
  \end{equation*}
  where the target is the probability of an epileptic state occurring in the $t$-th segment of channel $\mathbf{c}$ in the future segment set. 
\end{definition}

We emphasize here that our problem is defined on an individual patient.
A similar problem has been defined in \cite{tzallas2009epileptic}; however, their results were all based on EEG data, which are much simpler than SEEG data due to limited tracking of deep brain activities.
To the best of our knowledge, we are the first to formally propose and analyze the epileptic detection task on SEEG data.
We summarize important notations in Appendix~\ref{subsec:notations}.

%% file: 4_model.tex
\section{Proposed Model}~\label{sec:model}
\vspace{-0.1in}
\subsection{General Description}
\label{subsec:overview}

In real-world scenarios, doctors diagnose epilepsy patients step by step: 
whether the patient suffers from seizures, 
which brain regions are suspected epileptic areas, 
and the particular location in the brain that is directly causing the seizures (which might be surgically removed). 
Inspired by this, we propose a novel framework \model, which employs a hierarchical structure to jointly model epileptic waves and their diffusion process in three different levels, ranging from high to low: 
the patient, the brain-region, and the channel levels (with each channel corresponding to a particular location in the brain). 
While the problem under study focuses on the channel level, we find that higher levels provide a macroscopic view, which is helpful in capturing epileptic waves more precisely and improves the model's robustness to handle severe noise in SEEG data. 
Figure~\ref{pic:story}(b) illustrates the hierarchical structure of \model. 

Orthogonal to the three levels, \model consists of three major components: \textit{pre-training} for input SEEG data, \textit{graph diffusion} over time and a \textit{prediction} module.
We begin from the channel level to explain how these components work together. 
Given the massive SEEG data, with low epileptic wave rate, we first pre-train the representation of every segment to fully capture and leverage the patterns of normal brain waves by performing a self-supervised learning task (Section~\ref{subsec:ssl}).
To capture the dependency between each segment and the propagation of brain waves ignored by the pre-training component, we design the \textit{graph diffusion} component. 
Specifically, with the pre-trained representation taken as input, the \textit{graph diffusion} component incorporates a graph neural network along with a structure learning algorithm to explore how brain waves diffuse among channels (or brain regions in the higher level), and encodes these diffusion patterns into the segment representation (Section~\ref{subsec:gdm}). 
The obtained representation then can be directly piped to a classifier for \textit{prediction}. 
Similar procedures are followed at the brain-region and patient levels, where we use an aggregation strategy to obtain the brain-region/patient level representations from the channel/brain-region level (Section~\ref{subsec:multi}).
In summary, each component of the model is designed to capture every segment's own characteristics and its property of dynamic diffusion on the three different levels. 
In the remainder of this section, we introduce the details of each component.

\subsection{Model Pre-training}~\label{subsec:ssl}

Within the massive amount of SEEG records obtained from a patient lasting for days, there are often only seconds of epileptic waves. 
To handle this issue and fully leverage the massive (unlabeled) data, we propose a self-supervised learning (SSL) algorithm, Bidirectional Contrastive Predictive Coding (BCPC), for pre-training the representation of SEEG data. 

For a self-supervised algorithm, the key to learning discriminative representations is building an effective \textit{pretext task}~\cite{jing2020self}. 
By following Contrastive Predictive Coding (CPC)~\cite{oord2018representation}, 
the classical SSL model for time series, we set the pretext task as to predict the $P$ local features obtained from the encoder on the head and tail of the segment, with global contextual features acquired from the autoregression model based on the local features from the middle part. 
Different from CPC, we believe a bidirectional model will be more effective by enabling the contextual features to have the semantic information from both directions in the time dimension, which has also been illustrated in existing works such as BERT~\cite{devlin2018bert} and ELMo~\cite{peters2018deep}. 
Therefore, we design our pre-training model to predict the pretexts in both directions in a skip-gram fashion. 

\begin{figure}[t]
  \centering
  \includegraphics[width=\linewidth]{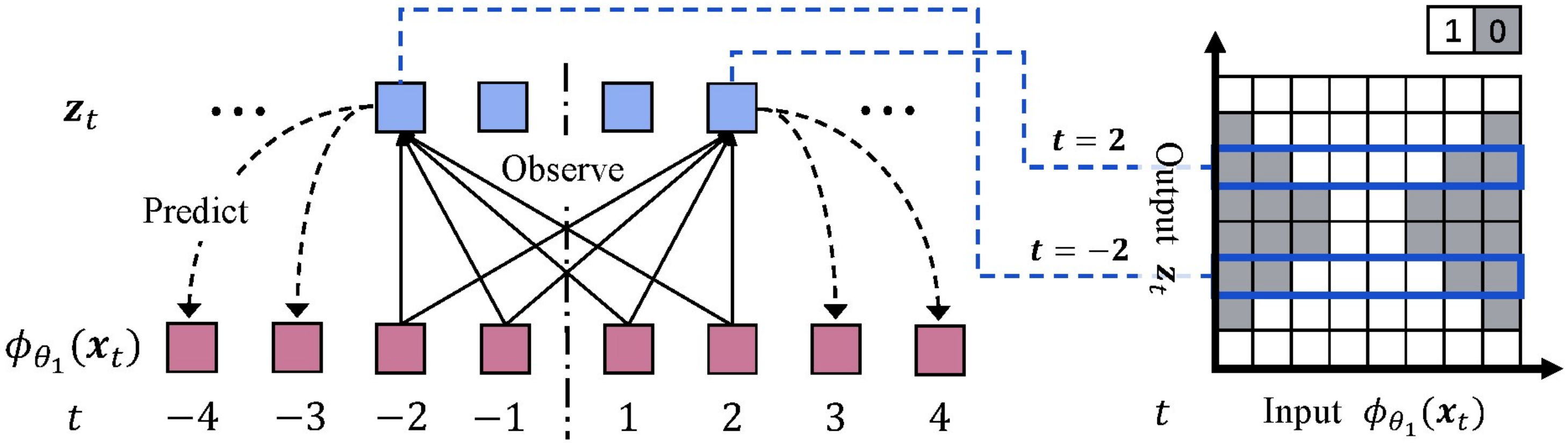}
  \caption{
  Schematic diagram of BCPC: 
  the left represents the pretext task; the right is the corresponding mask matrix. 
  }
  \label{pic:bcpc}
\end{figure}

Specifically, we use a pre-defined multi-layer CNN $\phi_{\theta_{1}}$ to embed the raw SEEG signals as \textit{local features} $\phi_{\theta_{1}}(x_{t})$. 
In order to ensure the model's ability to extract bidirectional information, we adopt a Transformer~\cite{vaswani2017attention} $\psi_{\theta_{2}}$ with a designed mask matrix as the autoregression model to obtain the \textit{global contextual features} $\boldsymbol{z}$. 
As Figure~\ref{pic:bcpc} shows, we divide a sequence into two equal half sub-sequences: the left-hand one is used to encode reverse direction information, while the right-hand one is for the forward direction. We set the center of the sequence as $0$ in the time coordinates, and assign positive (negative) subscripts to elements in the forward (backward) sequence.
With this definition, the $t$-th row in the mask matrix indicates the observable local features when constructing the global contextual features $z_t$, \emph{i.e.},  $\boldsymbol{z}_{t}=\psi_{\theta_{2}}(\phi_{\theta_{1}}(x_{t}), \ldots, \phi_{\theta_{1}}(x_{-t}))$. 
After that, based on the InfoNCE loss, we let $\boldsymbol{z}_{t}$ predict the unobserved local features that is $p$-step away from the global context, \emph{i.e.},   $\phi_{\theta_{1}}(x_{t+\text{sgn}(t) \cdot p})$.
Formally, we define our contrastive loss at time $t$ as
\begin{equation}
    \mathcal{L}_{t}=-\frac{1}{P} \sum_{p=1}^{P} \log
    \left[\frac{\text{Score}_{p} \left(\phi_{\theta_{1}}(x_{t+\text{sgn}(t) \cdot p}), \boldsymbol{z}_{t} \right)}{\sum_{\mathbf{n} \in \mathcal{N}_{t}} \text{Score}_{p} \left(\phi_{\theta_{1}}(\mathbf{n}), \boldsymbol{z}_{t} \right)}\right],
\end{equation}
where $\mathcal{N}_{t}$ denotes the random noise subset of the negative samples plus one positive sample, while $\text{sgn}(\cdot)$ is the signed function. 
A bilinear classifier is utilized for prediction.

\begin{figure*}[ht]
  \centering
  \includegraphics[width=\linewidth]{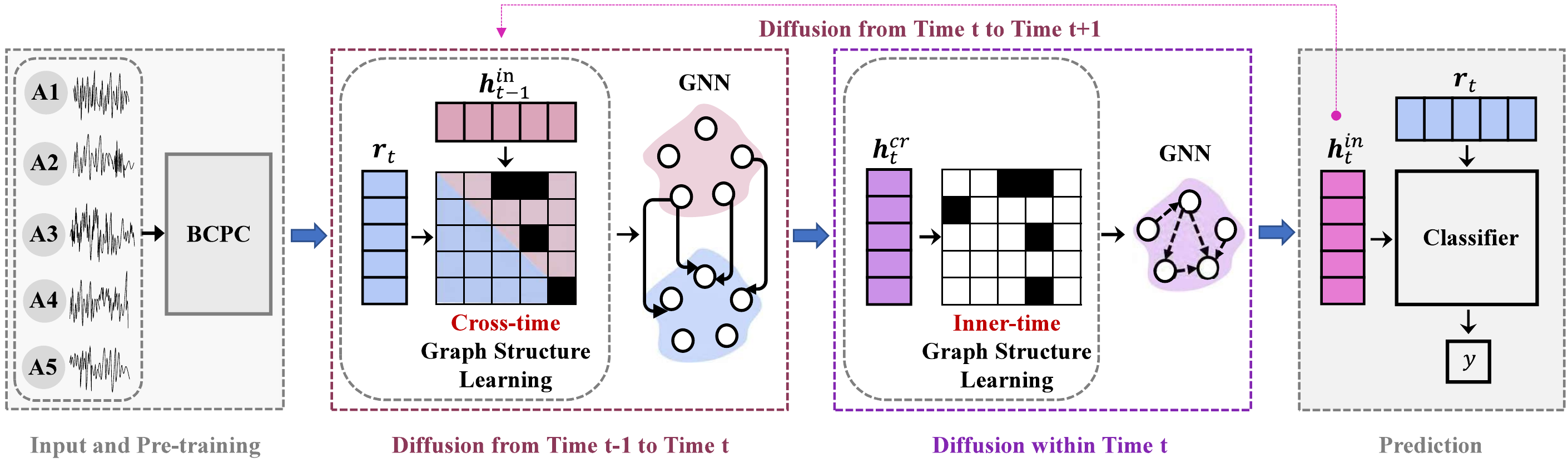}
  \caption{Architecture of \model at the channel level.
  \small
  The first part represents the process of obtaining segment representation $\boldsymbol{r}_t$ using the pre-trained BCPC. The second part describes the cross-time diffusion process from $t-1$ to $t$, in which the cross-time diffusion graph is first generated by the cross-time graph structure learning sub-module, and then a GNN is utilized on the learned graph. The third part corresponds to the inner-time diffusion process, which is based on a mechanism similar to the cross-time process. The outputs of the part will be utilized in the cross-time diffusion process of the next time step. The fourth part takes the input of the iteratively diffused representation $\boldsymbol{h}_t^{in}$ and combines it with the original representation $\boldsymbol{r}_t$ to predict whether an epileptic seizure occurs at time $t$.
  }
  \label{pic:gdm}
\end{figure*}

\subsection{Graph Diffusion Learning}~\label{subsec:gdm}

Although we have obtained the representation of every segment, interactions between channels and the underlying diffusion patterns have been ignored. 
We further propose a \textit{graph diffusion} component to explicitly model the diffusion process in the human brain. 
More specifically, we adopt the approach of alternating between graph structure learning and brain wave diffusion to achieve the target, in an end-to-end data-driven fashion.

\vpara{Graph structure learning.}
Overall, the challenge for modeling graph diffusion is that we do not know the underlying correlation and diffusion paths among the channels. 
We therefore propose to learn a graph structure where all the channels are the nodes. 
The key to tackling this problem is how to quantify the impact from one channel to another. 
Inspired by the phenomenon that in the brain's electrical activities, there universally exist \textit{traveling waves}~\cite{davis2020spontaneous}, which keep some characteristics such as shape and frequency during the propagation across different channels, we aim to propose our structure learning algorithm based on it. 

Under the assumption that similar channels are more likely to carry the same \textit{traveling waves} that propagate over time, we use the \textit{cosine} function as our metric to measure the correlation between pairs of channels. 
We leave the exploration of other similarity functions that might be more appropriate for time series data as future work.
Considering the asymmetric time delays of the traveling waves, we need to distinguish the direction of correlation. 
Therefore, we use a pair of source and target learnable weight parameters $\boldsymbol{W}_{1}$ and $\boldsymbol{W}_{2}$ to identify important features before the similarity computation. 
Formally, we obtain the score matrix $\mathcal{A}_{0}$ between every node pair from the cartesian product of source and target node sets $\boldsymbol{v}_{1} \times \boldsymbol{v}_{2}$ with node features $\boldsymbol{h}_{1}$ and $\boldsymbol{h}_{2}$ as follows:
\begin{equation}
    \mathcal{A}_{0}(i, j)=\cos{(\boldsymbol{W}_{1} \odot \boldsymbol{h}_{1}(i), \boldsymbol{W}_{2} \odot \boldsymbol{h}_{2}(j))},  ~\label{for:cosine}
\end{equation}
where $i \in \boldsymbol{v}_{1}$, $j \in \boldsymbol{v}_{2}$ and $\odot$ denotes the Hadamard product. 
Notice that, $\mathcal{A}_{0}$ is very likely to represent a dense graph.
To maintain the sparsity of graph structure that works of neuroscience suggest~\cite{yu2017connectivity, achard2007efficiency}, and remove insignificant and spurious connections caused by low frequency fluctuation or physiological noises, we filter unnecessary edges by using a threshold-based filter function  $F_{\theta}(\cdot)$ with a tunable hyper-parameter $\theta$ as follows: 
\begin{equation}
    F_{\theta}(x)=\left\{
    \begin{aligned}
    &x, \qquad x \geq \theta; \\ 
    &0, \qquad x < \theta.
    \end{aligned}\right.   ~\label{for:filter}
\end{equation}
Finally, the graph structure $\mathcal{A}(i, j) = F_{\theta}(\mathcal{A}_{0}(i, j))$ is obtained.

\vpara{Brain wave diffusion.}
The constructed graph $\mathcal{A}$ represents the relative correlation between channels. 
The greater the edge weight, the more possible diffusion will occur. 
We aim to trace the diffusion along the constructed graph to enhance the representation of the traveling waves. 
During seizures, more rapid and significant propagation of spike-and-wave discharges will appear~\cite{proix2018predicting}, which implies more distinguishable representations after propagation. 
We therefore adopt GNNs to model the brain wave diffusion process due to their natural message-passing ability on a graph.
For simplicity and clear performance attribution, we use the standard GCN model in this work and take the one-layer directed GCN~\cite{yun2019graph} as an example to describe the process.
Specifically, given the graph $\mathcal{G}=((\boldsymbol{v}_{1}, \boldsymbol{v}_{2}), \mathcal{A}, (\boldsymbol{h}_{1}, \boldsymbol{h}_{2}))$, we obtain the representation of the target node $j \in \boldsymbol{v}_{2}$ after diffusion as follows:
\begin{equation}
    \boldsymbol{h}_{2}^{\prime}(j) = \sigma \left(\frac{\boldsymbol{h}_{2}(j) + \sum_{i \in \boldsymbol{v}_{1}} \mathcal{A}(i, j) \cdot \boldsymbol{h}_{1}(i)}{1 + \sum_{i \in \boldsymbol{v}_{1}} \mathcal{A}(i, j)} \Theta \right),     ~\label{for:gnn}
\end{equation}
where $\Theta$ is the learnable linear transformation matrix for directed GCN and $\sigma(\cdot)$ denotes the ReLU activation function.

\vpara{Putting it all together.}
Owing to the epileptic waves lasting longer or shorter, \model learns two types of diffusion processes. 
Concretely, \textit{cross-time diffusion} naturally models the propagation of longer epileptic waves between two consecutive time segments.
Meanwhile, fast signal spreading within the same time segments of each channel are captured by \textit{inner-time diffusion}.
Based on the structure learning and brain wave diffusion process, we formulate~\eqref{for:cosine}-\eqref{for:gnn} as a function $\mathcal{F}_{\boldsymbol{W}_{1}, \boldsymbol{W}_{2}, \Theta}(\boldsymbol{v}_{1}, \boldsymbol{v}_{2}, \boldsymbol{h}_{1}, \boldsymbol{h}_{2}, \theta)$.
We will then use the notation to introduce the graph diffusion component involving the two types of diffusion. 

Graph diffusion component executes the two diffusion steps in the order of cross-time followed by inner-time.
Given the representations obtained from the ($t-1$)-th inner-time diffusion process $\boldsymbol{h}^{\text{in}}_{t-1}$, we derive the cross-time diffusion at the $t$-th time segment:
\begin{equation}~\label{eq:cross}
    \boldsymbol{h}^{\text{cr}}_{t} = \mathcal{F}_{\boldsymbol{W}^{\text{cr}}_{1}, \boldsymbol{W}^{\text{cr}}_{2}, \Theta^{\text{cr}}}(\mathbf{C}_{t-1}, \mathbf{C}_{t},  \boldsymbol{h}^{\text{in}}_{t-1}, \boldsymbol{r}_{t}, \theta^{\text{cr}}),
\end{equation}
\noindent where both the source and target node sets consist of all channels,
\emph{i.e.}, $\mathbf{C}_{t}=\mathbf{C}$ for $t=1,\dots,|S|$. 
Here we use the subscript only to emphasize the time index. 
Essentially, the cross-time diffusion models the impact from the representation of last segments ($\boldsymbol{h}^{\text{in}}_{t-1}$) to that of current segments ($\boldsymbol{r}_{t}$), which is obtained from Section~\ref{subsec:ssl}.

Following the cross-time diffusion at the $t$-th time segment, with a slight difference in the formula, the inner-time process traces the correlations among the representation $\boldsymbol{h}^{\text{cr}}_{t}$ of segments within the same time:
\begin{equation}~\label{eq:inner}
    \boldsymbol{h}^{\text{in}}_{t} = \mathcal{F}_{\boldsymbol{W}^{\text{in}}_{1}, \boldsymbol{W}^{\text{in}}_{2}, \Theta^{\text{in}}}(\mathbf{C}_{t}, \mathbf{C}_{t},  \boldsymbol{h}^{\text{cr}}_{t},  \boldsymbol{h}^{\text{cr}}_{t}, \theta^{\text{in}}).
\end{equation}
The output representation $\boldsymbol{h}^{\text{in}}_{t}$ obtained after inner-time diffusion will then be involved in the next cross-time diffusion and, eventually, to the prediction stage. 
It is worth mentioning that the first cross-time diffusion has no historical representations of previous time segments. To handle this, we define a virtual node set $\boldsymbol{v}_{0}$ and virtual representation $\boldsymbol{h}_{0}$ to construct an empty graph and diffuse on it.

As shown above, Eq.\eqref{eq:cross} and Eq.\eqref{eq:inner} separately models the two diffusion patterns, and the two steps are alternately connected. 
Until now, we have acquired the representations after the graph diffusion component for further classification. Moreover, we also process the segments in the reverse time direction independently to obtain the reverse representations after inner-time diffusion $\boldsymbol{h}^{\text{in}}_{-t}$ at the $t$-th time segment.

\subsection{Hierarchical Predictions}~\label{subsec:multi}

Given the representations obtained from the graph diffusion component, we concatenate $\boldsymbol{h}^{\text{in}}_{t}$, $\boldsymbol{h}^{\text{in}}_{-t}$ and $\boldsymbol{r}_{t}$ together so as to obtain the predicted probability of seizures $\hat{y}_{t}$ through a discriminator $D$ that implemented by a two-layer MLP.
The objective function for epileptic wave detection of the channel is then defined as the binary cross-entropy:
\begin{equation}
\label{eq:object_ch}
    \mathcal{L}_{\text{ch}} = -\sum_{t=1}^{|S|} \sum_{\mathbf{c} \in \mathbf{C}} \left[ y^{S}_{t,\mathbf{c}} \log \hat{y}_{t,\mathbf{c}} + (1 - y^{S}_{t,\mathbf{c}}) \log (1 - \hat{y}_{t,\mathbf{c}}) \right].
\end{equation}

However, the diffusion process for the channel level alone has limited horizons and lacks a more macroscopic perception.
As Section~\ref{subsec:overview} inspires, the synthesized information of the channel, brain-region, and patient level will be considered simultaneously to facilitate more accurate diagnoses.
We describe the hierarchical task design combining these three levels in more details below.

\vpara{Hierarchical label construction.}
The labels are originally situated at the channel level. Following the reverse logic order of doctor's diagnosis, we mark a brain region as epileptic if at least one of the channels in it is epileptic. Similarly, the patient is regarded as being in a seizure state if at least one brain region is abnormal. 
More formally, given the $t$-th segment labels with all channels $y_{t}^S$ (see Section~\ref{sec:pre}), we first divide the channels into different brain regions through the mapping $b(\cdot)$. Then for each brain region $\mathbf{b} \in \mathbf{B}$, we assign the segment label as follows: 
\begin{equation}
    y_{t,\mathbf{b}}^{\text{brain}} = \max_{\mathbf{c}: b(\mathbf{c})=\mathbf{b}} y_{t,\mathbf{c}}^S.
\end{equation}
In a similar fashion, we can also construct the patient-level's segment labels based on labels of brain regions: 
\begin{equation}
    y_{t}^{\text{patient}} = \max_{\mathbf{b} \in \mathbf{B}} y_{t,\mathbf{b}}^{\text{brain}}.
\end{equation}

\vpara{High-level representation learning.}
We aggregate representations from the lower levels to obtain representations in the higher levels. 
While there are many available choices for permutation-invariant pooling, we adopt max pooling to aggregate the representations, because this approach is more aligned with the hierarchical label construction process and we need the features that contribute most to the epileptic state in the lower levels. 
Formally, taking the brain region representations $\boldsymbol{r}^{\text{brain}}$ as an example, we use the element-wise max-pooling method as follows: 
\begin{equation}
    \boldsymbol{r}^{\text{brain}}_{t,\mathbf{b}}(i) = \max_{c:b(c)=\mathbf{b}} \boldsymbol{r}_{t,c}(i),\qquad i=1,\dots,d,
\end{equation}
where $\mathbf{b} \in \mathbf{B}$ and $d$ denotes the dimension of the representations.

After the pooling operation, the graph diffusion component is implemented on the high-level representations to obtain the respective objectives, \emph{i.e.}, $\mathcal{L}_{\text{br}}$ and $\mathcal{L}_{\text{pa}}$, each of which is defined as the binary cross-entropy similar to Eq.\eqref{eq:object_ch}.
Considering that the completely independent parameters cannot guarantee a consistent optimization direction, we therefore let the three levels share the same parameter sets in the graph diffusion component and the discriminator $D$ in order to align their representation spaces.

Finally, we jointly optimize the three tasks in different levels.
With the guidance of labeled data, we optimize the proposed model via back propagation and learn the representation spaces for epileptic wave detection. 
Through the hierarchical task framework, \model is expected to aggregate accurate information at higher levels, and in turn feed it back to the lower levels, resisting the data noise and improving the model performance.

%% file: 5_experiment.tex
\section{Experiments}~\label{sec:exp}

\vspace{-0.08in}
\subsection{Dataset}

\vpara{Data collection.}
The SEEG dataset used in our experiment is provided by a first-class hospital. The dataset statistics is summarized in Table~\ref{tab:dataset_statistics} (Appendix~\ref{subsec:ds}).
More specifically, for a patient suffering from epilepsy, \textit{4} to \textit{10} invasive electrodes with \textit{52} to \textit{126} channels are used for recording \textit{256}Hz to \textit{1024}Hz SEEG signals; these figures vary from patient to patient. Notably, as SEEG signals are collected with high-frequency across multiple channels, our dataset is massive. 
In total, we have collected \textit{526} hours of SEEG signals with \textit{769}GB.
Although each prediction is patient-specific, to validate the generalizability and stability of our model, we repeat all experiments on multiple patients.
Based on our dataset, two professional neurosurgeons helped us to label the epileptic waves. 
We regard all time points within an epileptic wave as positive samples and the remainder as negative ones. 
The positive sample ratio of a single patient in our dataset is around \textit{0.003} on average, which is extremely imbalanced. 
The dataset will be released after further cleansing and scrutinization.

\vpara{Preprocessing.}
In order to conduct the self-supervised learning task (Section~\ref{subsec:ssl}), for each patient, we randomly sample {10000} normal segments with a window length of {1} second. 
Ninety percent of the sampled segments are then used for training, with the remainder used for validation. 
As for the epileptic wave detection task, for each patient, 
we first obtain {13300} segments to train our model ({85$\%$} for training and {15$\%$} for validation). 
For the testing, to explore the performance of different models under datasets with different positive sample ratios, we sample three test sets, each of which includes {1140}, {9690} and {95190} segments for each patient, with positive-negative sample ratio of \textit{1:5}, \textit{1:50} and \textit{1:500} respectively.

\subsection{Experimental Setup}
First, we train a model for each patient independently and evaluate the performance. Then we repeat the experiments on all patients and obtain the average results.
Owing to different numbers of recording channels in different brain regions for different patients, generalization between patients is difficult, and we will leave this issue as future work. Therefore, we only conduct experiments on a single patient.
Considering the advantage conferred by the hierarchical task framework, we validate the effectiveness of our proposed \model at both the channel and patient levels. 
As for the baselines, to the best of our knowledge, no existing model can handle these two levels of tasks at the same time.
Therefore, for each task, we adopt task-specific baselines.
Specifically, for the \textit{channel-level epileptic wave detection task},
we compare \model with several univariate time series classification models including TSF~\cite{deng2013time}, STSF~\cite{cabello20fast}, MiniRocket~\cite{dempster2021minirocket}, WEASEL~\cite{schafer2017fast}, LSTM-FCN~\cite{karim2017lstm} and TS-TCC~\cite{ijcai2021-324}. 
For each baseline, we train an independent model for every single channel and obtain the average results on all channels for one patient.
As for the \textit{patient-level epileptic wave detection task}, we use the following multivariate time series classification models as baselines: EEGNet~\cite{lawhern2018eegnet}, TapNet~\cite{zhang2020tapnet}, MLSTM-FCN~\cite{karim2019multivariate} and NS~\cite{franceschi2019unsupervised}.
We provide details of these baselines, the evaluation metrics and the hyperparameter analysis in the Appendix. 
These baselines are designed to deal with raw time series data rather than representation space, so we do not use representations pre-trained by BCPC as input to these models.

\begin{table*}[ht]
  \caption{The average performance of epileptic wave detection tasks at the channel and patient levels. Test datasets with different positive-negative sample ratios are used for evaluation.}
  \setlength\tabcolsep{3.3pt}
  \centering
  \label{tab:main_exp}
  \begin{tabular}{l|l|ccccc|ccccc|ccccc}
    \toprule
    \multicolumn{2}{c|}{} & \multicolumn{5}{c|}{$1:5$} & \multicolumn{5}{c|}{$1:50$} & \multicolumn{5}{c}{$1:500$} \\
    \cmidrule{3-17}
    \multicolumn{2}{c|}{\multirow{-2.5}{*}{\diagbox[width=8em,trim=l,height=2.5\line]{\raisebox{0.3ex}{\textbf{Models}}}{\raisebox{-0.3ex}{\textbf{Ratio}}}}} & Pre. & Rec. & $F_1$ & $F_2$ & AUC & Pre. & Rec. & $F_1$ & $F_2$ & AUC & Pre. & Rec. & $F_1$ & $F_2$ & AUC \\
    \midrule
    \multirow{7}*{\shortstack{Channel \\ Level}} & TSF & 51.22 & 45.33 & 44.02 & 44.16 & 83.47 & 16.99 & 46.62 & 21.92 & 29.71 & 85.00 & 2.42 & 36.52 & 3.96 & 7.52 & 79.66 \\
    & STSF & 53.51 & 57.65 & 51.79 & 54.26 & 86.75 & 18.21 & 56.98 & 24.69 & 34.48 & 87.80 & 2.61 & 49.73 & 4.63 & 9.22 & 84.07 \\
    & MiniRocket & 55.68 & 56.84 & 53.55 & 54.84 & 87.07 & 21.82 & 55.30 & 28.12 & 36.85 & 87.47 & 4.02 & 48.93 & 6.83 & 12.42 & 84.10 \\
    & WEASEL & 50.38 & 43.92 & 43.84 & 43.24 & 80.93 & 17.91 & 43.61 & 22.44 & 29.11 & 82.21 & 2.79 & 36.83 & 4.84 & 9.05 & 77.50 \\
    & LSTM-FCN & 36.32 & 33.85 & 30.35 & 30.99 & 73.41 & 10.73 & 35.64 & 13.91 & 19.65 & 75.68 & 1.68 & 30.12 & 2.79 & 5.38 & 70.88 \\
    & TS-TCC & 56.20 & 53.48 & 50.81 & 51.58 & 85.57 & 21.60 & 52.59 & 26.87 & 34.79 & 86.33 & 3.93 & 46.63 & 6.40 & 11.41 & 82.49 \\
    \cmidrule{2-17}
    & {\model} & \textbf{68.19} & \textbf{62.71} & \textbf{61.90} & \textbf{61.59} & \textbf{95.22} & \textbf{42.74} & \textbf{58.21} & \textbf{45.29} & \textbf{50.36} & \textbf{94.80} & \textbf{16.87} & \textbf{51.69} & \textbf{21.88} & \textbf{30.06} & \textbf{92.19} \\
    \midrule
    \multirow{6}*{\shortstack{Patient \\ Level}} & EEGNet & 55.24 & 64.62 & 57.37 & 60.74 & 84.81 & 13.20 & 64.02 & 21.07 & 33.56 & 83.32 & 1.38 & 54.73 & 2.66 & 6.04 & 78.25 \\
    & TapNet & 71.48 & 57.93 & 63.37 & 59.88 & 91.73 & 30.55 & 61.89 & 38.33 & 47.56 & 90.15 & \multicolumn{5}{c}{\rule[2pt]{0.1\textwidth}{0.1pt}\ OOM\ \rule[2pt]{0.1\textwidth}{0.1pt}} \\
    & MLSTM-FCN & 68.68 & 70.96 & 66.77 & 68.70 & 89.60 & 28.00 & 73.32 & 34.38 & 46.14 & 90.69 & 4.39 & 62.25 & 7.43 & 13.47 & 85.50 \\
    & NS & 57.68 & 54.99 & 55.80 & 55.19 & 80.76 & 14.70 & 54.13 & 22.52 & 33.70 & 80.17 & 1.67 & 44.88 & 3.20 & 7.06 & 73.24 \\
    \cmidrule{2-17}
    & {\model} & \textbf{79.61} & \textbf{79.08} & \textbf{76.69} & \textbf{76.87} & \textbf{94.92} & \textbf{40.31} & \textbf{80.53} & \textbf{49.37} & \textbf{60.30} & \textbf{93.86} & \textbf{12.36} & \textbf{72.58} & \textbf{18.57} & \textbf{28.85} & \textbf{91.93} \\
  \bottomrule
\end{tabular}
\end{table*}

\subsection{Experimental Results}

The average performance over all patients of different methods are presented in Table~\ref{tab:main_exp}. 
Overall, \model outperforms all baselines on every evaluation indicator in both channel and patient levels. 

\vpara{Results in channel-level.}
In the channel-level task, \model improves \textit{12.31$\%$}, \textit{36.66$\%$}, \textit{142.03$\%$} in terms of $F_2$ on the test datasets with positive-negative sample ratio of 1:5, 1:50 and 1:500 respectively. 
In particular, as the labels become more and more imbalanced, the performance of baselines drops rapidly, while \model keeps a relatively much better performance than baselines. The increasing ratio of performance improvement implies our model has the ability to handle more imbalanced data, which is more aligned with the practical clinical scenarios.
Compared with univariate baselines, \model takes the advantage of learning how the epileptic waves diffuse across channels.
For example, two channels A1 and A2 have a low correlation in the normal state, \emph{i.e.}, the diffusion of the normal brain waves between them is very weak.
However, during the seizures, their correlation increases significantly, which is reflected in the larger edge weight in the diffusion graph learned by our model.
When \model determines that A1 contains seizures, it can further infer the epileptic status of A2 with more certainty through the diffusion process along with the learned graph structure.
This makes the prediction result of A2 more reliable compared with the situation only considering A2 alone.

\vpara{Results in patient-level.}
At the patient level, \model also improves the performance by {\textit{{50.95$\%$}}} in terms of $F_2$ in average on the three test datasets. 
The results show the superiority of our hierarchical task design to make more accurate predictions through utilizing more refined information at lower levels.
In more detail, \model can offer some evidence of the specific local (channel- and brain-region-level) information for the global system (patient), which benefits the detection task in patient-level. For example, being aware of which particular brain region has epileptic waves increases the model's confidence when inferring whether the patient is suffering from epilepsy.

\subsection{Ablation Study}

In this section, we conduct ablation experiments to verify the effectiveness of each major component in our model. 
More specifically, we remove each of the following components from our model to see how it influences the performance respectively: 
pre-training (\model-BCPC), inner-time diffusion step (\model-Inner), cross-time diffusion step (\model-Cross), hierarchical task framework (\model-Multi) and graph diffusion component (\model-graph). 

\begin{table}[ht]
  \caption{Results of ablation study.}
  \label{tab:ablation_exp}
  \small
  \setlength\tabcolsep{4.8pt}
  \begin{tabular}{l|l|ccccc}
    \toprule
    \multicolumn{2}{c|}{\diagbox[width=9em,trim=l]{\textbf{Models}}{\textbf{Metrics}}} & Pre. & Rec. & $F_1$ & $F_2$& AUC \\
    \midrule
    \multirow{7}*{\shortstack{Channel \\ Task}} 
    &  \model-BCPC & 3.86 & 33.84 & 5.26 & 8.42 & 80.12 \\
    &  \model-Graph & 3.26 & 20.15 & 4.00 & 6.07 & 71.61 \\
    &  \model-Inner & 11.43 & 38.58 & 13.76 & 17.99 & 91.14 \\
    &  \model-Cross & 4.78 & 41.40 & 7.56 & 12.81 & 85.05 \\
    &  \model-Multi & 11.80 & 44.19 & 14.70 & 19.68 & 88.81 \\
    \cmidrule{2-7}
    & {\model} & \textbf{16.87} & \textbf{51.69} & \textbf{21.88} & \textbf{30.06} & \textbf{92.19} \\
    \midrule
    \multirow{4}*{\shortstack{Patient \\ Task}} 
    &  \model-BCPC & 1.22 & 49.84 & 2.32 & 5.09 & 81.71 \\
    &  \model-Inner & 5.27 & 62.24 & 9.20 & 17.08 & 88.12 \\
    &  \model-Cross & 3.60 & 56.32 & 6.54 & 13.08 & 86.79 \\
    \cmidrule{2-7}
    & {\model} & \textbf{12.36} & \textbf{72.58} & \textbf{18.57} & \textbf{28.85} & \textbf{91.93}\\
  \bottomrule
\end{tabular}
\end{table}

We report the evaluation results of the ablation experiments on the test dataset with 1:500 positive-negative sample ratio in Table~\ref{tab:ablation_exp}. It can be observed that \model achieves the best performance to all ablated model versions in all metrics, which demonstrates the effectiveness of each component in our model design. 
For \model-BCPC, the striking drop in performance indicates the powerful representation ability of BCPC. 
Comparison with \model-Graph, which obtains the representations from BCPC and feeds them into MLP directly, reveals that our model achieves superior performance (improves more than \textit{390$\%$} in terms of $F_2$). 
It suggests the significance of modeling the diffusion process.

\subsection{Case Study}

\begin{figure}[t]
  \centering
  \includegraphics[width=\linewidth]{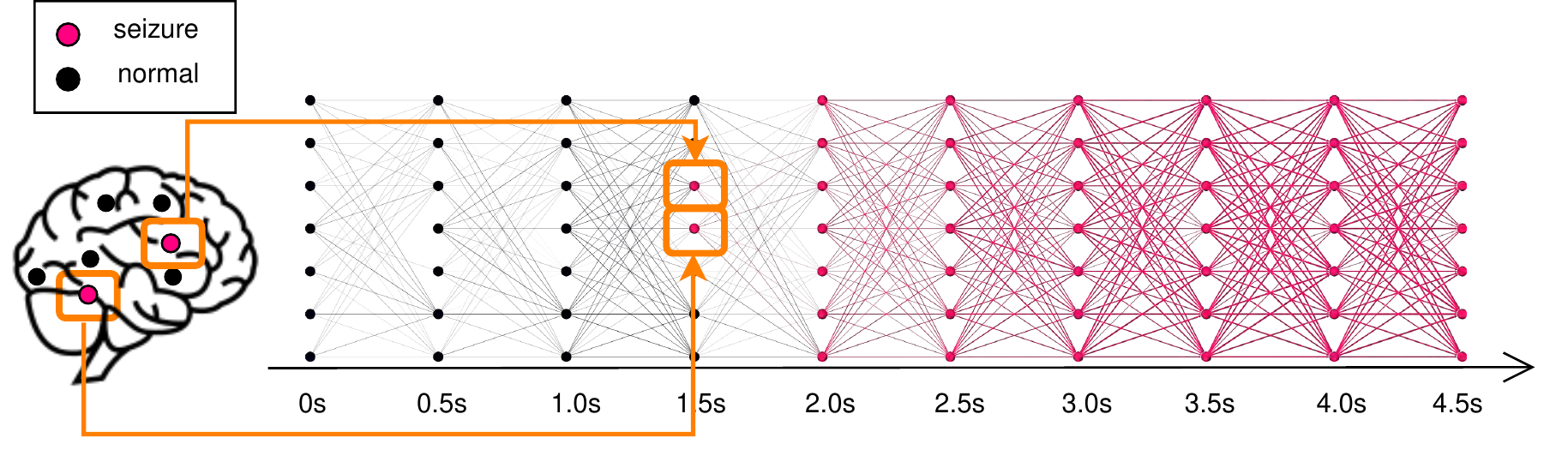}
  \caption{A case of cross-time diffusion. \small 
  The horizontal direction is arranged in chronological order with 10 contiguous segments at intervals of 0.5s, while the 7 different brain regions in patient P-4 are expanded in the vertical direction. 
  The black nodes denote the normal state of the corresponding brain regions predicted by \model, and red ones indicate the predicted epileptic brain regions.
  }
  \label{pic:cross_case}
\end{figure}

We at last present a case study to illustrate how cross diffusion works. 
As shown in Figure~\ref{pic:cross_case}, 10 contiguous time segments are presented with a corresponding learned cross diffusion graph structure.
We can see that at time \textit{1.5}s epileptic waves appear in two brain regions for the first time, and then diffuse to other brain regions. The edge weights are lower in the normal state, while edges with high weights are widely observed when epileptic waves appear in multiple brain regions. This phenomenon is consistent with some domain knowledge in the field of neuroscience. More specifically, dramatic changes of brain connectivity in SEEG can be tracked during seizures~\cite{bartolomei2017defining}.
During normal status, most brain regions have relatively weak connections. When a seizure occurs, brain regions gradually form a resonance, which means that strong connections between brain regions will be observed. This phenomenon increases the credibility of our epilepsy diffusion graph. Moreover, if a seizure occurs in the two brain regions as first time predicted by our model, there is a higher probability that SOZ will be located in these two brain regions, which can assist the doctors with diagnosis in clinical practice.

\subsection{Online System}~\label{subsec:os}

We have deployed \model into an online system, where doctors could upload the SEEG data of patients. Then \model will detect the epileptic waves from the data and present the results in the visualization panel. Therefore, doctors can review the predicted epileptic waves very quickly to obtain the basic seizure patterns of patients. This will save much time for doctors to develop further treatment plans.

\begin{figure}[t]
  \centering
  \includegraphics[width=\linewidth]{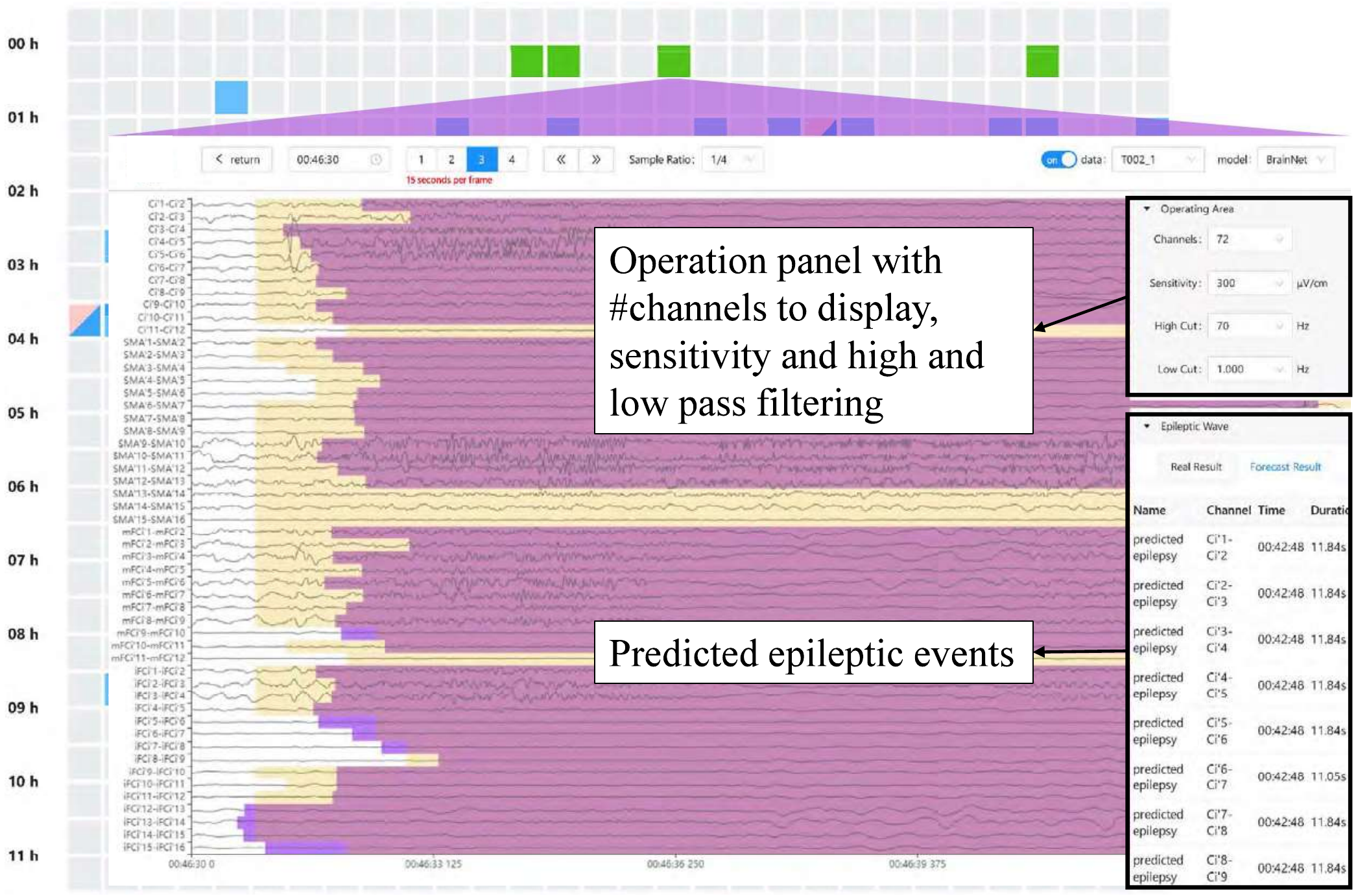}
  \caption{
  Demonstration of the online system.
  \normalsize
  }
  \label{pic:online}
\end{figure}

Figure~\ref{pic:online} shows the screenshot of the system interface. 
The top part of Figure~\ref{pic:online} is the profile page of \textit{patient overview}.
We show the overview of a 12-hour patient file after being reviewed by doctors.
Each square with different colors denotes a 1-minute data segment. The gray squares $(\vcenter{\hbox{\includegraphics[width=2.1ex,height=2.1ex]{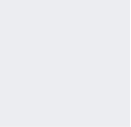}}})$ denote that no epileptic waves exist, while the green ones $(\vcenter{\hbox{\includegraphics[width=2.1ex,height=2.1ex]{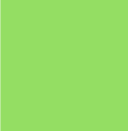}}})$, the blue ones $(\vcenter{\hbox{\includegraphics[width=2.1ex,height=2.1ex]{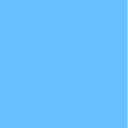}}})$ and the red ones $(\vcenter{\hbox{\includegraphics[width=2.1ex,height=2.1ex]{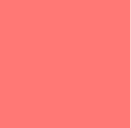}}})$ represent correct, wrong and missing predictions of our model respectively.
Doctors can see the detailed SEEG data that may contain a seizure onset predicted by our model by clicking one square.
The bottom part of Figure~\ref{pic:online} shows the prediction results of epileptic wave detection of the clicked square. 
The top toolbar is used to change the presented time period. The data operation panel and epileptic wave events can be found on the right.
In the center of the page, the purple part represents the real epileptic waves labels, if any, offered by doctors and the yellow part is our model's prediction. 
As shown in the figure, the predictions of our model match the actual seizures well.

%% file: 6_related.tex
\section{Related Work}~\label{sec:related}

Epilepsy is a disorder of the brain that can be detected using SEEG signals. Features of the SEEG are patient-specific in nature and vary largely from one person to another.
Much work has been conducted over the past few decades in an attempt to design an automated system that can analyze and detect seizures and predict them before their occurrence, so that required measures can be taken to record them.
Numerous works have been conducted by researchers to understand epilepsy and the characteristics of brain activity that accompanies an epilepsy attack so as to detect and predict the onset of the seizure. The earliest of these studies can be traced back to a 1982 work by \textit{Gotman}~\cite{gotman1982automatic}, who developed patient non-specific detectors.
Seizures consist of various kinds of wave-forms, such as spikes, sharp waves, sleep spindles, and periods~\cite{fergus2016machine}. Accurate patient-specific detectors were designed using SVM classifiers in~\cite{fotiadis2015handbook}. Sensitivity and false detection rate were used as the standard for measuring performance by researchers in~\cite{tzallas2012automated}. There are limitations in automatically detecting and predicting seizures~\cite{furbass2015prospective}.

Using traditional machine learning methods to detect epileptic waves requires feature engineering, which in turn requires large amounts of domain knowledge and is also time-consuming. As an end-to-end method, deep learning is widely employed to perform this task. \textit{Roy et al.} combine a one-dimensional convolutional layer and Gated Recurrent Unit (GRU) to perform epileptic wave detection~\cite{roy2018deep}. \textit{Hisham Daoud et al.} design a deep convolutional autoencoder architecture that pre-trains the model in an unsupervised manner. After pre-training, the trained encoder is connected to a Bidirectional Long Short-Term Memory (Bi-LSTM) Network for classification~\cite{daoud2019efficient}. \textit{Lawhern et al.} propose a model named EEGNet, which is composed of two-dimensional convolution layers and pooling layers and was designed for EEG-based brain-computer interfaces; it can also be used to perform epileptic wave detection based on EEG~\cite{lawhern2018eegnet}.

%% file: 7_conclusion.tex
\section{Conclusion}~\label{sec:conclusion}
In this paper, we proposed a novel model, \model, to learn the diffusion graph via a hierarchical framework for epileptic wave detection. 
Drawing on domain knowledge from neuroscience, to detect epileptic waves more accurately, we study the \textit{epileptogenic network}, learn the underlying dynamic diffusion graph structures, and model the cross-time and inner-time diffusion patterns at the channel, brain region, and patient levels.
Moreover, by conducting experiments on a large-scale real-world dataset, we demonstrate the effectiveness of our proposed model on detecting epileptic waves and modeling the diffusion process.
Furthermore, we have deployed \model into an online system to assist the diagnosis of epilepsy, where doctors could upload the SEEG data and obtain the epileptic waves identified by our model (see details in Section~\ref{subsec:os}). 
In the future, motivated by our positive results on both the accuracy of epileptic wave detection and interpretability of learned epileptogenic graphs, it would be intriguing to further validate our framework across more epileptic patients and study its real-world clinical deployment. 

\vpara{Acknowledgment.}
This work is supported by NSFC (No.62176233), the National Key Research and Development Project of China (No.2018AAA0101900) and the Fundamental Research Funds for the Central Universities.

%% file: 8_appendix.tex
\clearpage

\section{SUPPLEMENT}~\label{sec:appendix}

In the supplement, we provide the details not expanded in the text above, including the notations used in our model, dataset statistics, evaluation metrics of experiments, details of our baseline methods, and results of the hyperparameter analysis.

\subsection{Notations}~\label{subsec:notations}

The notations used in our model are summarized in Table~\ref{tab:notations}.

\begin{table}[h]
  \caption{Summary of important notations.}
    \label{tab:notations}
    \begin{tabular}{cl}
    \toprule
    \textbf{Notation} & \textbf{Description} \\
    \midrule
    $\mathbf{B}, \mathbf{C}$ & The set of brain regions and channels\\
    $b(\cdot)$ & The function mapping each channel to the \\
    & corresponding brain region\\
    $t, \mathbf{c}, \mathbf{b}$ & The index of segments, channels and brain\\
    & regions\\
    $s_{t, \mathbf{c}}$ & The data of t-th segment of channel $\mathbf{c}$\\
    $y^{S}_{t, \mathbf{c}}$ & The label of t-th segment of channel $\mathbf{c}$\\
    $r_{t}$ & The representation acquired by the pre-training\\
    &BCPC and a linear transformation at the $t$-th\\
    &segment \\
    $h^{\text{in}}_{t}$ & The representation of the $t$-th segment after\\
    & inner-time diffusion process\\
    $h^{\text{cr}}_{t}$ & The representation of the $t$-th segment after\\
    & cross-time diffusion process\\
    \bottomrule
  \end{tabular}
\end{table}

\subsection{Dataset Statistics}~\label{subsec:ds}

The SEEG data related to each patient is placed in Table~\ref{tab:dataset_statistics}, including recording time, sampling frequency, the number of electrodes inserted into the brain, the number of channels contained in the electrodes, the epileptic wave ratios and the total samples for each patient. During the experimental phase, the channels actually used by each patient removed those unrelated to seizures and bad channels that doctors told us.
\begin{table}[t]
  \caption{Statistics of SEEG data corresponding to each patient in our dataset. Notice that we detect epileptic waves on segment level and channel level, and obtain 55,961,080 samples in total (5,596,108 samples for each patient on average).}
  \setlength\tabcolsep{2.7pt}
  \small
  \label{tab:dataset_statistics}
  \begin{tabular}{lrrrrrr}
    \toprule
    \multirow{2}{*}{Patients} & \multirow{2}{*}{\shortstack{Time \\ (Hours)}} & \multirow{2}{*}{\shortstack{Sample \\ frequency}} & \multirow{2}{*}{\shortstack{\# Elec- \\ trodes}} & \multirow{2}{*}{\shortstack{\# Chan- \\ nels}} & \multirow{2}{*}{\shortstack{Epileptic \\ wave ratio}} & 
    \multirow{2}{*}{\# Samples}\\
    \\
    \midrule
    P-1 & 72 & 1000Hz & 10 & 126 & 0.003 & 8,113,760 \\
    P-2 & 21 & 1000Hz & 4 & 52 & 0.004 & 3,460,280 \\
    P-3 & 114 & 1000Hz & 10 & 126 & 0.001 & 9,664,920 \\
    P-4 & 56 & 256Hz & 4 & 52 & 0.001 & 4,295,520 \\
    P-5 & 36 & 512Hz & 5 & 63 & 0.002 & 3,818,240 \\
    P-6 & 6 & 512Hz & 5 & 69 & 0.004 & 3,698,920 \\
    P-7 & 24 & 512Hz & 7 & 89 & 0.009 & 6,801,240 \\
    P-8 & 36 & 1024Hz & 4 & 52 & 0.001 & 1,789,800 \\
    P-9 & 24 & 512Hz & 7 & 93 & 0.0002 & 4,176,200 \\
    P-10 & 137 & 1000Hz & 8 & 110 & 0.0006 & 10,142,200 \\
  \bottomrule
\end{tabular}
\end{table}

\subsection{Evaluation Metrics}~\label{subsec:em}

In our experiments, we use the following metrics to evaluate the models: precision, recall, two F-measures (\emph{i.e.} $F_1$ and $F_2$) and AUC. Notice that we do not consider the accuracy metric because of the extremely imbalanced sample ratio. The F-measure is a metric defined as the weighted harmonic mean of precision and recall, with the following mathematical formulation:
\begin{equation*}
    F_{\beta} = \frac{(1+\beta^{2}) \times precision \times recall}{\beta^{2} \times precision + recall}.
\end{equation*}

In our experiments, $F_2$ is more favored than $F_1$, as in clinical context, missing any epileptic waves is costly.

\subsection{Baselines}~\label{subsec:baseline}

We give details of our baseline methods used in two levels of tasks. 

\vpara{Channel-level epileptic wave detection.} 
We compare \model with several univariate time series classification models. 
For each baseline, we train an independent model for every single channel and obtain the average results on all channels for one patient. 

\begin{itemize}[leftmargin=*]
    \item Time Series Forest (TSF)~\cite{deng2013time}: This is a tree-ensemble method for time series classification. The features employed in TSF are basic time-series features, including the mean, standard variance, and slope of a segment.
    
    \item Supervised Time Series Forest (STSF)~\cite{cabello20fast}: This is an interval-based tree model that adopts a top-down approach to search for relevant subseries in three different time series representations before training any tree classifier.
    
    \item MiniRocket~\cite{dempster2021minirocket}: This approach reformulates Rocket~\cite{dempster2020rocket} into a fast version maintaining essentially the same accuracy.
    
    \item WEASEL~\cite{schafer2017fast}: This approach transforms time series into feature vectors using a sliding window approach. The vectors are then analyzed through a machine learning classifier.

    \item LSTM-FCN~\cite{karim2017lstm}: This is used for univariate time series classification consisting of an LSTM layer and CNN layer.
    
    \item TS-TCC~\cite{ijcai2021-324}: This is a contrastive method that learns time series representation through two different views generated by weak and strong augmentations of the original segments.

\end{itemize}

\vpara{Patient-level epileptic wave detection.}
We use the following multivariate time series classification models as baselines:

\begin{itemize}[leftmargin=*]
    \item EEGNet~\cite{lawhern2018eegnet}: This is a model proposed in the field of neuroscience. Specifically, it is a compact convolutional neural network for EEG-based brain-computer interfaces. 
    
    \item TapNet~\cite{zhang2020tapnet}: An attentional prototype network,  which takes the strengths of both traditional and deep learning based approaches to perform multivariate time series classification.
    
    \item MLSTM-FCN~\cite{karim2019multivariate}: This is a deep learning framework consisting of an LSTM layer and stacked CNN layer, along with a Squeeze-and-Excitation block for multivariate time series classification.

    \item NS~\cite{franceschi2019unsupervised}: This approach uses an unsupervised method employing time-based negative sampling to learn embeddings. SVM is then applied to perform the final classification.

\end{itemize}

Considering the trade-off between efficiency and effectiveness, we run TapNet 200 epochs with learning rate 10 times larger than the default. For NS, we run 100 epochs with default learning rate. All other baselines are initialized with the same parameters suggested in their respective papers. 

\subsection{Hyperparameter Analysis}~\label{subsec:ha}

\begin{figure}[t]
    \subfigure[Inner threshold]{
    \includegraphics[width=0.47\linewidth]{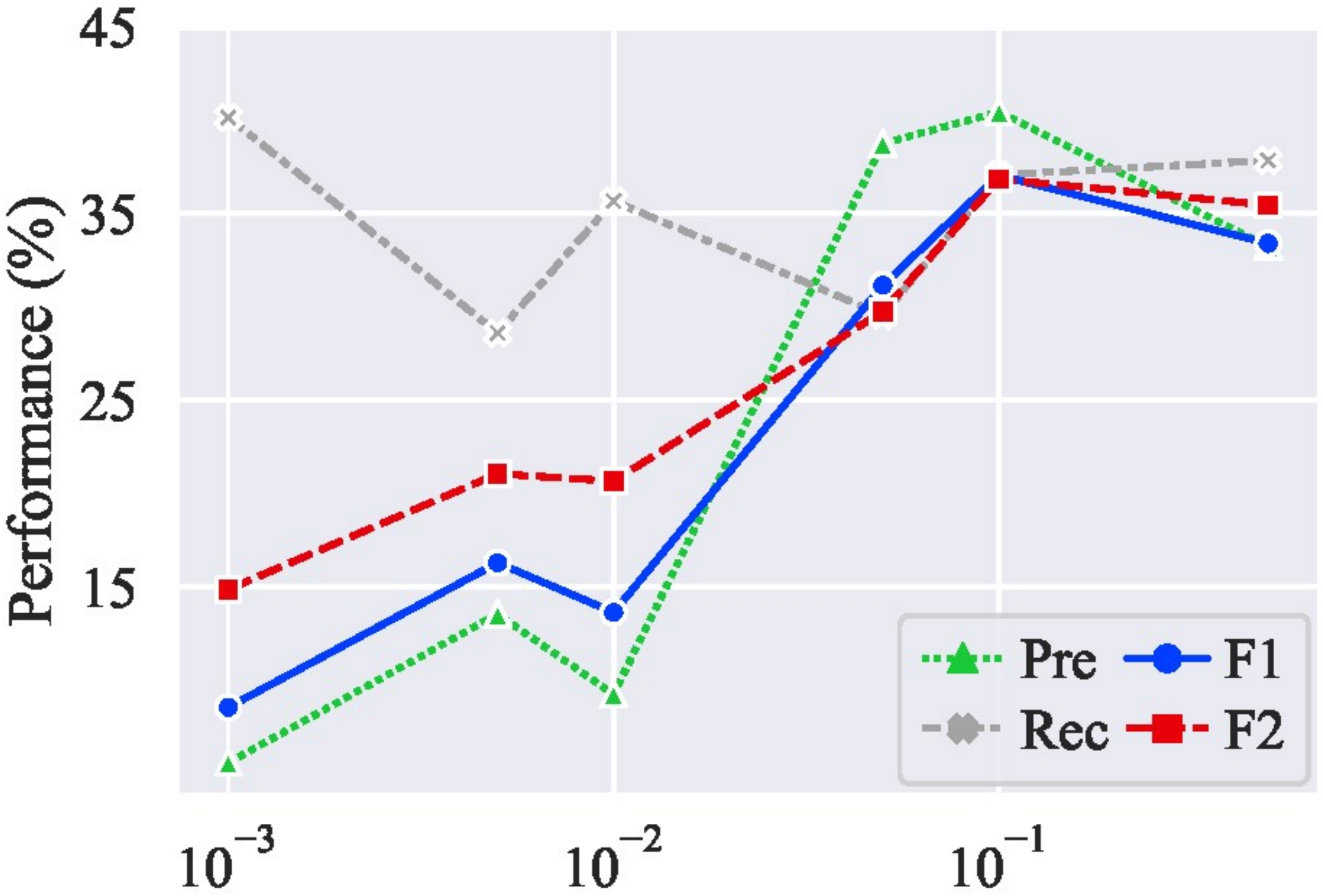}
    }
    \subfigure[Cross threshold]{
    \includegraphics[width=0.47\linewidth]{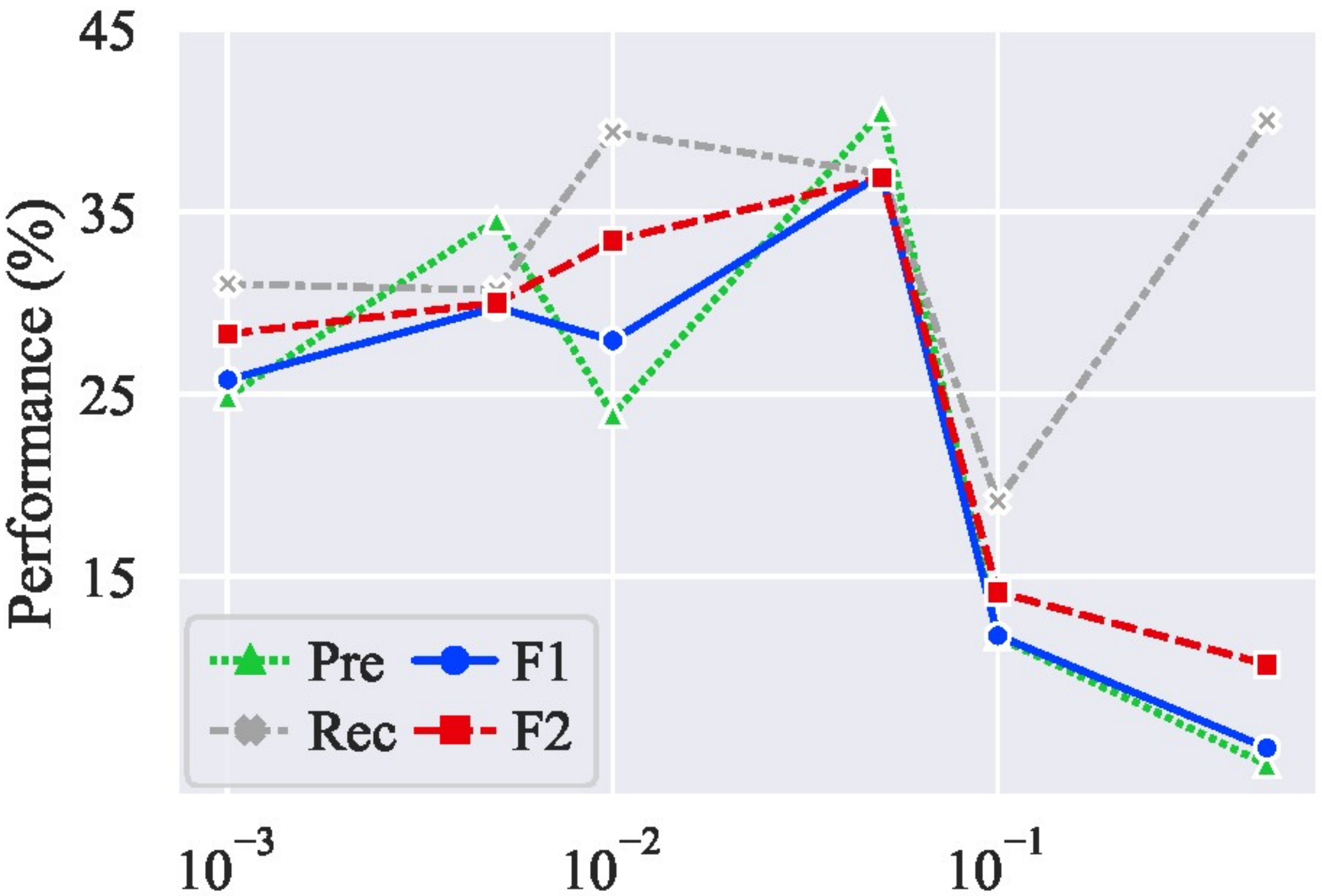}
    }
    \caption{The hyperparameter analysis of the thresholds in the inner-time diffusion (a) and the cross-time diffusion (b).}
    \label{pic:threshold}
\end{figure}

We analyze the influence of the hyperparameters-the two thresholds in the graph diffusion component-on the performance (Figure~\ref{pic:threshold}). 
Taking the patient P-6 on the test dataset with $1:50$ sample ratio as an example, whose best inner-time and cross-time thresholds are 0.1 and 0.05 respectively, the results show that the performance of \model reaches the best results near the two optimal thresholds and degrades on both sides of the optimal values. 
This can be interpreted as more noisy correlations will be introduced along with the more dense diffusion graph structure if we set a relatively low threshold. On the contrary, a too high threshold will also harm the performance because some representative diffusion patterns might be ignored.